# Characteristics of Cost Overruns for Dutch Transport Infrastructure Projects and the Importance of the Decision to Build and Project Phases

By

Chantal C. Cantarelli, Eric J. E. Molin, Bert van Wee, and Bent Flyvbjerg

## Introduction

Transportation infrastructures are expensive and often involve large cost overruns. Projects become more expensive than was initially estimated and additional budget is required. Consequently, as the total budget for infrastructure investments is generally fixed, the budget to cover the costs of other projects is insufficient. Cost overruns therefore not only result in financial consequences for the project under consideration but also ultimately in fewer infrastructure projects being realised than planned. The problem of cost overruns is even more disturbing considering the fact that "cost escalation has not decreased over the past 70 years" (Flyvbjerg et al., 2003b).

The problem of cost overruns is thus severe and persistent and probably affects every country investing in transport infrastructure. The question, however, is to what extent. The study by Flyvbjerg et al. (2002, 2003a, 2004) can be considered the leading piece of research into cost overruns because of the large number of projects included, the variety of project types, the long time period and the wide geographical coverage. However, the results of this study do not necessarily apply for individual countries. For example, the conclusions for Europe do not necessarily apply to each individual European country.

Several studies on cost overruns have focused on individual countries. Table 1 gives an overview of these studies on cost overruns, their geographical area, the frequency and the magnitude of cost overruns. The study of Flyvbjerg et al. (2003a) is included in the table for reasons of comparison. Note that these studies measure cost overruns slightly differently, which is explained a few lines below the table.

**Table 1 Frequency and magnitude of cost overruns found in literature [a]**

| Study | Geographical area | Frequency cost overrun (%) | Magnitude of cost overrun | | | | | | | |
|---|---|---|---|---|---|---|---|---|---|---|
| | | | *Road* | | *Rail* | | *Fixed Links* | | *Other* | |
| | | | % | N | % | N | % | N | % | N |
| Merewitz (1973) | US | 79 | 26 | 49 | 54 | 17 | | | | |
| Morris (1990) | India | | | | 164 | 23 | | | 4 | 10 |
| Pickrell (1990, 1992)[b] | US | 88 | | | 61 | 8 | | | | |
| Auditor General (1994)[c] | Sweden | | 86 | 8 | 17 | 7 | | | | |
| Nijkamp and Ubbels (1999) | Netherlands, Finland | 75 | | | | | | | 0-20 | 8 |
| Bordat et al. (2004) | US | 55 | 5 | 2668[d] | | | | | | |
| Odeck (2004) | Norway | 52 | 8 | 620 | | | | | | |
| Dantata et al. (2006) | US | 81 | | | 30 | 16 | | | | |
| Ellis et al. (2007) | US | | 9 | 3130 | | | | | | |
| Lee (2008)[e] | South Korea | 95 | 11 | 138 | 48 | 16 | | | | |
| | | | | | | | | | | |
| Flyvbjerg et al. (2003a) | World | 86 | 20 | 167 | 41 | 58 | 34 | 33 | | |

[a] In which: %: the percentage cost overrun and N: the number of projects with cost overruns
[b] In van Wee (2007)
[c] In Odeck (2004)
[d] Projects include: Road and bridge construction and rehabilitation projects; maintenance projects, with road maintenance and resurfacing contracts; Traffic and traffic maintenance contracts
[e] In Siemiatycki (2009)

All of these studies show that cost overruns are more common than cost underruns, with frequencies ranging between 52% and 95%. Conversely, the magnitude of the cost overruns differs between the studies. As for the differences in average cost overruns between studies, the following main explanations can be given. First of all, the main reason for the differences in the average cost overrun between studies is the difference in the use of nominal and real prices (Flyvbjerg, 2007). Secondly, the way data are handled can explain the differences in the extent of cost overruns between studies (see for a more extensive elaboration Flyvbjerg et al., 2003b). Studies use a different moments for the year of decision to build and the year of completion as the basis for the estimated and actual costs, and hence the extent of the cost overruns differs. Thirdly, differences can also be related to the variation in sample size. If the sample size is small, outliers may have a large influence on the results. Fourthly, the differences can be explained by the differences in the geographical area that is covered (different economies), and the project types that are included (different project dynamics and complexity).

Table 1 shows furthermore that the studies by Nijkamp and Ubbels (1999), Odeck (2004), Bordat et al. (2004) and Ellis et al. (2007) found rather small cost overruns – up to 20% – whereas Morris (1990) and the Auditor General of Sweden (in Odeck) found enormous cost overruns of 164% and 86% respectively. Due to the probability that there are large differences in cost overruns among countries and across time, cost overruns in country specific studies might be quite different (either lower or higher) from the average cost overruns in the worldwide study. However, the abovementioned findings do not support this expectation; some country specific studies have smaller average cost overruns and others higher average cost overruns compared to the worldwide study by Flyvbjerg et al. Ideally country specific studies should have the same methodology as this worldwide study in order to compare its results. Without a general tendency in country specific studies towards lower or higher average cost overruns than in the worldwide study, it remains difficult to make inferences about individual countries based on the worldwide database. This therefore supports the need for further research into country specific cost performance of transport infrastructure projects.

The objective of this research is to determine the characteristics of cost overruns in Dutch large-scale transport infrastructure projects. This concerns the frequency and the magnitude of cost overruns and whether cost estimates have improved over time. This research was financed by the Dutch Ministry of Infrastructure and the Environment and the Netherlands was therefore chosen as the country under scrutiny.

We apply the same methodology as used for the worldwide research regarding project types, the way data and prices are used. A database of 78 Dutch large-scale transportation infrastructure projects was created and statistical analyses were used to determine the frequency and magnitude of cost overruns in the Netherlands and to examine whether cost estimates have become more accurate over time.

In the literature on cost overruns (see table 1) hardly any attention is given to the project phases (with the exception of Odeck (2004)). Until now literature has focused on identifying cost overruns but the moment when projects are most vulnerable to cost increases has not been studied. This is however of the utmost importance because it could improve our understanding of cost overruns considerably. Whether cost increases are incremental over different project phases or extreme in certain project phases will help to distinguish between different explanations of cost overruns. It is essential to look more closely when cost increases occur to actually be able to deal with them. A second objective of this research is therefore to investigate whether projects are more vulnerable to cost increases during different project phases and if so which phase this concerns. A distinction was made between 2 phases (data did not allow to distinguish between more phases): 1. the *pre-construction phase* (the period between the formal decision to build and the start of construction) and 2. the *construction phase* (the period between the start of construction and the start of operation (opening)). By addressing project phases in relation to project performance this paper fills a gap in literature on cost overruns and project management.

The structure of the paper is as follows. Section 2 describes the project selection, data collection, representativeness of the database, and the main methodological issues. Section 3 presents the cost performance in the Netherlands, focussing on the characteristics of cost overruns. In section 4 a comparison is made between the cost overruns in the pre-construction phase and in the construction phase. Finally, section 5 discusses the main conclusions and section 6 presents several areas for further research.

## Project Selection, Data Collection and Methodology

### Definition Large-Scale Projects

*Large-scale* projects are often defined as major infrastructure projects that cost more than US$1 billion (Flyvbjerg et al., 2003a). However, past studies have often included a wider range of projects, both smaller sized projects costing several million dollars and large-scale projects (see e.g. Flyvbjerg et al., 2003b, where the smallest project cost US$ 1.5 million, and Odeck, 2004, which included projects costing less than 15 million NOK ~ US$ 12.3 million). In addition to the size of the project in terms of costs, large-scale projects attract a high level of public attention or political interest because of substantial direct and indirect impacts on the community, environment, and budgets (FHWA in Capka, 2004). Therefore, the definition of a large-scale project can also depend on the context, that is, the size of the project in relation to the size of the city (or country). Based on project size, their impact and context, projects that cost more than about € 20 million are considered large-scale projects in the Netherlands[1]. Regarding *transport infrastructures*, we adopt the definition of van Wee (2007): "Transport infrastructures include roads, rail lines, channels, (extensions of) airports and harbours, bridges and tunnels. Of these projects the 'hardware' is considered, excluding 'software'; projects that are not related to the construction of infrastructure but are related to policies of deregulations, liberalization, privatization, and so forth". In line with previous studies, the project types that are included in this research are road, rail, tunnel and bridge projects.

### Project Selection and Data Collection

All large-scale transport infrastructure projects in the Netherlands that were completed after the year 1980 were selected. Projects completed before this year are excluded because the data were expected to be difficult to come by.

Data were collected from a variety of sources, i.e. interviews with former project leaders and project teams; archives research at the Ministry of Infrastructure and the Environment; RWS[2] Direction Large Projects and RWS Direction Zuid-Holland; internet search; and the MIRT reports. The MIRT (*Meerjarenprogramma Infrastructuur, Ruimte en Transport*, translated as the Multi-year programme for infrastructure, spatial planning and transport[3]) was a valuable source of information for both identifying large-scale road and rail projects and collecting data. The MIRT is the implementation programme related to the policy of 'mobility and water'. It is funded by the Infrastructure Fund of the Ministry of Infrastructure and the Environment. The MIRT[4] includes all infrastructure projects in the Netherlands. For this research the programmes for the years 1984-2010[5] were accessed.

Based on the MIRT, 70 road and 39 rail projects were identified (one rail project falls out of the period 1984-2010 but is included because data for this project was readily available from the research of Flyvbjerg et al. (2003a)). Of these projects, 34 projects (23 road and 11 rail) were rejected for reasons of limited data availability, and 12 projects (10 road and 2 rail) were rejected because the data was invalid.

To identify tunnels we used an international database and gallery for structures – Structurae – (http://en.structurae.de) and to identify bridges we used the database of the National Bridge Foundation (NBF) (http://bruggenstichting.nl). However, neither of these databases includes data on costs and in order to select the *large-scale* projects, the length was used as a surrogate criterion for project size. Larger projects have a greater impact on the community, environment, and budgets and they require more effort to fit into the landscape, not only due to the development density but also for aesthetic reasons. Since the level of effort differs between bridges and tunnels, the definition of a large-scale project based on the project length differs as well. Bridge projects, for example, have a larger influence on the visual hindrance compared to tunnel projects and hence the minimum length of a large-scale project is less for bridge projects. The minimum length for projects is based on construction cost indices and was set at 500 meters for tunnels and 200 meters for bridges (http://www.bouwkostenkompas.nl)[6]. Data on fixed link projects was collected by means of interviews and archive research. In total, 27 tunnel and 25 bridge projects were identified. For 38 of these projects data for crucial variables (e.g. costs or decision to build) were missing, so these projects could not be included in the database. To summarise, the database consists of 78 projects.

As may have become clear, the resulting database does not include all projects due to incompleteness of information, which may be regarded as non-response (thus not due to sampling mechanisms, because these were not applied). However, in line with previous international research in this field that also includes projects based on data availability, the database is considered to be a sample. In this research also non-significant differences will be reported because we are also interested in a complete description of the project performance of the specific projects in the database.

**Representativeness**

For road and rail projects, the database is fairly representative. Data collection was based on the MIRT and was not dependent on retrieving information directly from project managers. Therefore bias, in the sense that managers may have an interest in whether or not data is provided and whether this is accurate or presented in a favourable light, probably does not play a role. For the fixed link projects, data was partly collected by means of interviews with project managers involving the risk of bias mentioned above. However, the average cost overruns for fixed link projects for which data was collected by means of interviews is not statistically different from the average cost overruns for fixed link projects for which data was collected by means of documentation research. (t=-1.414, p=0.188, N=15, independent sample t-test). The results need reservation because of the small number of projects.

**Methodology**[7]

The two most important data variables in this research are the estimated and actual costs. Cost overrun is measured as actual out-turn costs minus estimated costs expressed as a percentage of the estimated costs. Actual costs are defined as real, accounted construction costs determined at the time of project completion. Estimated costs are defined as budgeted or forecasted construction costs determined at the Time of formal Decision to build (ToD). This is also called the "decision date", "the time of the decision to proceed," the "go-decision" (Flyvbjerg et al., 2003a). At that moment, cost estimates were often available as data for decision-makers to make an informed decision.

*Estimated costs* are the costs at the ToD. In line with Flyvbjerg et al. (2003a), when the costs are not available at the ToD, the nearest available reliable figure for estimated costs is used as a proxy. This is typically a later estimate, which is often more accurate, and therefore leads to lower cost overruns. By investigating the cases with complete information, it was estimated that the cost overruns presented in this study are about 1% lower because of this assumption. We did not correct cost estimates using this figure because it is based on many assumptions and only concerns a small deviation. Note that usually contingency costs are included in the estimated costs. The cost overruns as referred to in this study are therefore in excess of these contingency costs.

The *actual construction costs* are the costs at the year of completion (year operations begun). If the actual costs are unknown at the time of project completion, the most reliable later figure for actual costs is used (i.e. from a year later than the opening year), if available. If unavailable, an earlier figure for actual costs was used (i.e., from a year before the opening year), but only if at least 90% of the budget was spent at this time, i.e., the project was at least 90% complete in financial terms. The cost overruns presented in this study are about 0.8% higher because of this assumption. For the same reason as given for estimated costs, we did not use this percentage as a correction factor.

All costs were converted to 1995 prices using the appropriate historical and sectoral indices for discounting in the Netherlands to correct for inflation (similar to Flyvbjerg et al. 2003a). Based on expert opinions[8] the most appropriate indices were determined, which included the GWW index, an index by ProRail for rail projects and the CROW index[9]. Research on cost overruns typically presents costs without VAT. VAT is, therefore, also excluded in the costs for the projects in this research. In adjusting the costs for VAT, the difference between a low and high tariff as well as changes of the tariff over the years is taken into account. The methodology of data collection and the calculation of cost overruns was approved by two independent authorities from the Ministry of Infrastructure and the Environment, i.e. RWS and KiM.

## Characteristics of Cost Overruns

### Magnitude of Cost Overruns

Figure 1 shows a histogram with the distribution of cost overruns for all Dutch transportation infrastructure projects in the database.

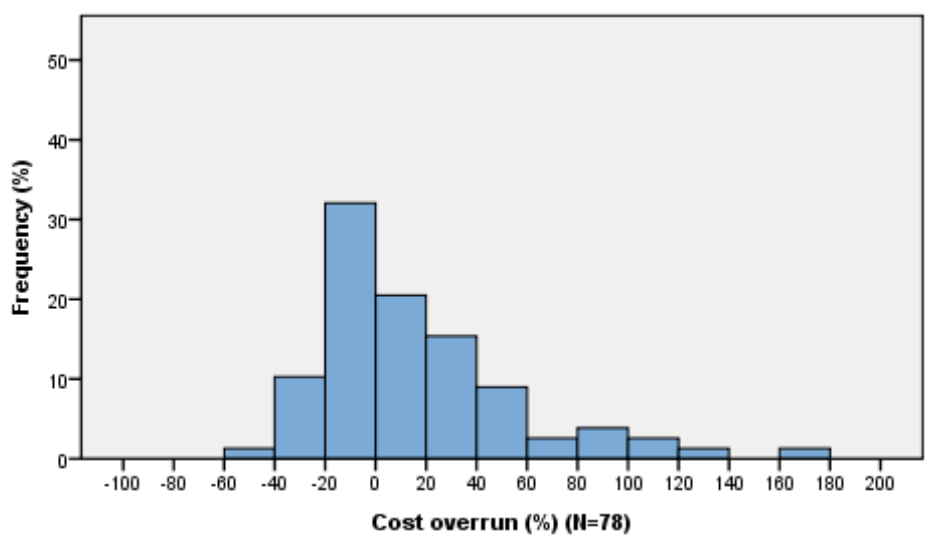

**Figure 1 Distribution of cost overruns in Dutch transportation infrastructure projects**

The histogram shows a large spread around zero indicating that the errors in forecasting costs are various and large. Furthermore, there is an asymmetrical distribution around zero, implying that errors in overestimating costs are different in size to errors in underestimating costs.

The specific statistics are as follows (figures rounded off to one decimal)[11]:

- The range of cost overrun is -40.3% to 164.0%
- The average cost overrun is 16.5%.
- The standard deviation is 40.0, indicating a rather large variation of the individual cost overruns around the mean.

Figure 1 shows two striking features in the distribution of cost overruns. First of all, there is one project, The Tweede Heinenoordtunnel, with an extremely large cost overrun of 164.0%. It was the first tunnel in the Netherlands to be bored, and the additional complexity involved with this construction method can partly explain the cost overruns. If this project is excluded, the total average cost overrun decreases to 14.6% (SD=36.5).

Secondly, a large number of projects (32%) have cost underruns in the category -20% to 0%. Considering this group of projects in more detail, the results indicate that this group does not differ from the other projects in the database regarding project type. However, the group does differ in terms of project size (expressed in line with standard convention by estimated costs); the projects are considerably smaller with €119 million compared to the other projects with an average size of €317 million (p=0.235, independent sample t-test).

**Frequency of Cost Overruns**

The main findings regarding the frequency of cost overruns are as follows:

- In 55% of the projects, actual costs are larger compared to estimated costs (resulting in cost overruns) whereas in 44% of the projects, actual costs are lower compared to estimated costs (resulting in cost underruns).
- Projects with cost overruns are as common as projects with cost underruns[10] (p=0.428, binominal test).
- Projects with cost overruns have an average overrun of 41.3% (SD=38.1). Projects with cost underruns have on average an underrun of 13.9% (SD=10.5) (Mann-Whitney U =0.000, p=0.000, Mann-Whitney U-test).

**Cost Performance Over Time**

In order to consider the project performance in terms of costs over time we consider time by the year of completion and the year of formal decision to build. According to Flyvbjerg et al. (2003b) "it is better to use year of decision to build rather than year of completion; the latter includes length of implementation phase, which has an influence on cost escalation, causing confounding". Data on the year of completion is however more evident and hence more reliable. We therefore consider both time variables (see figures 2 and 3). One project is excluded from the analysis as it was completed in 1970 whereas the other projects were all completed in the period 1991-2009.

Figure 2 does not give reason to assume a relation between the year of completion and cost overruns. Based on a regression analysis, we conclude that there is indeed no effect between both variables (F=0.002, p=0.964). For the relation between the year of decision to build and cost overruns (Figure 3) we also conclude that there is no statistical significant effect (F=2.486, p=0.119).

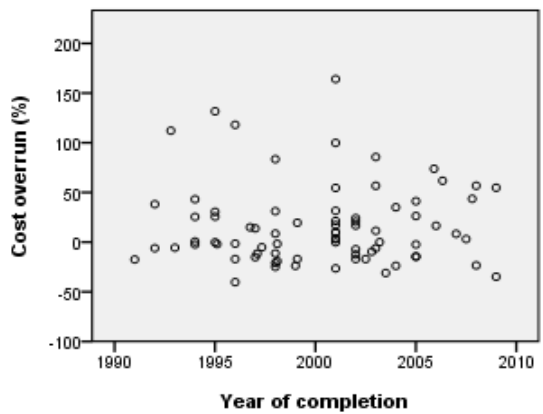

**Figure 2 Cost overruns over time (year of completion)**

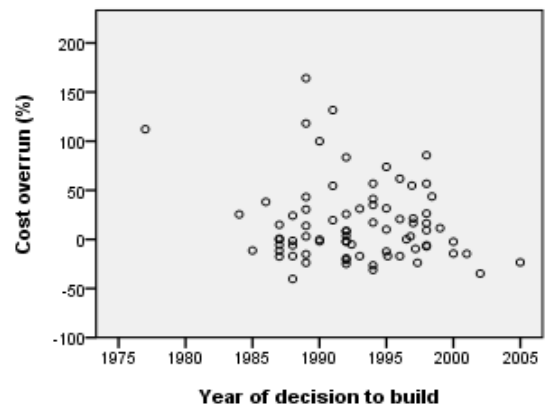

**Figure 3 Cost overruns over time (year of decision to build)**

Data showed that cost estimates have not improved over time. We therefore rule out technical explanations as the main reason for cost overruns in Dutch projects. These consider cost overruns the result of technical errors such as inadequate forecasting techniques. If technical explanations were the main cause of cost overruns, cost estimates should have improved over time since better methods have become available. Better methods would – all other factors remaining constant – result in better cost estimates, reducing the importance of the factors relevant for technical explanations for cost overruns. These findings are in line with the international research (Flyvbjerg et al. (2003a; 2003b).

In search for a possible explanation for the lack of improvement of cost estimates over time, the large construction fraud that was committed during the tendering for numerous governmental projects in the Netherlands could be considered. This construction fraud came to light in the year 2000, after which projects were set under stricter management. Therefore, if projects that were decided upon before 2000 have higher average cost overruns, construction fraud may be an explanation for the large and consistent cost overruns over time. It turns out that the average cost overrun for this group of projects (18.8%) is indeed significantly larger than the average cost overrun for projects that were decided upon from 2000 onwards (-18.0%) (t=2.013, p=0.048). However, the number of projects completed is considerably lower (7%) than the number of projects completed before the year 2000 (93%).

## Cost Overruns during Different Project Phases

This section discusses cost overruns during the project development. Note that actual cost overruns can only take place once the project is completed, but for reasons of simplicity we stick with this term. Hereby two project phases are distinguished: the pre-construction and the construction phase.

The pre-construction phase is the period between the formal decision to build (ToD) and the start of construction. The construction phase is the period between the start of construction and the year of completion. Only those projects for which data on the essential variables are available are included.

We excluded three projects for which data on the year of construction start were unavailable and fourteen projects for which data on the estimated costs at the time of construction start were unavailable. In some cases, construction started in the same year or even before the formal decision to build was made. These projects do not have an explicit pre-construction phase. The construction phase is then equal to the implementation phase and consequently the cost overrun in the construction phase is the same as the overall cost overrun. This would give a distorted picture regarding the phase in which the largest cost overruns take place and these projects were therefore not included in this analysis (24 projects).

To investigate the cost overruns during different phases, data from 37 projects were used. There is no systematic bias regarding these projects with respect to cost overruns compared to the other projects (t=0.483, p=0.630, independent sample t-test).

The cost overrun in the pre-construction phase is measured as the estimated costs at the start of construction minus the estimated costs at the ToD expressed as a percentage of the estimated costs at the ToD. The cost overrun in the construction phase is measured as the actual out-turn costs minus the estimated costs at the start of construction expressed as a percentage of the estimated costs at the start of the construction. Let us consider the project development consisting of three key moments: $T_0$: formal decision to build (ToD), $T_1$: start of construction, and $T_2$: project opening. Consequently, the estimated costs at these moments can be referred to as follows: $C_0$, $C_1$, and $C_2$ respectively, resulting in the following formulas for cost overruns:

$$\Delta C_{\text{pre-construction phase}} = \frac{C_1 - C_0}{C_0} \times 100\%$$

$$\Delta C_{\text{construction phase}} = \frac{C_2 - C_1}{C_1} \times 100\%$$

Figure 4 shows the distribution of cost overruns in the pre-construction phase.

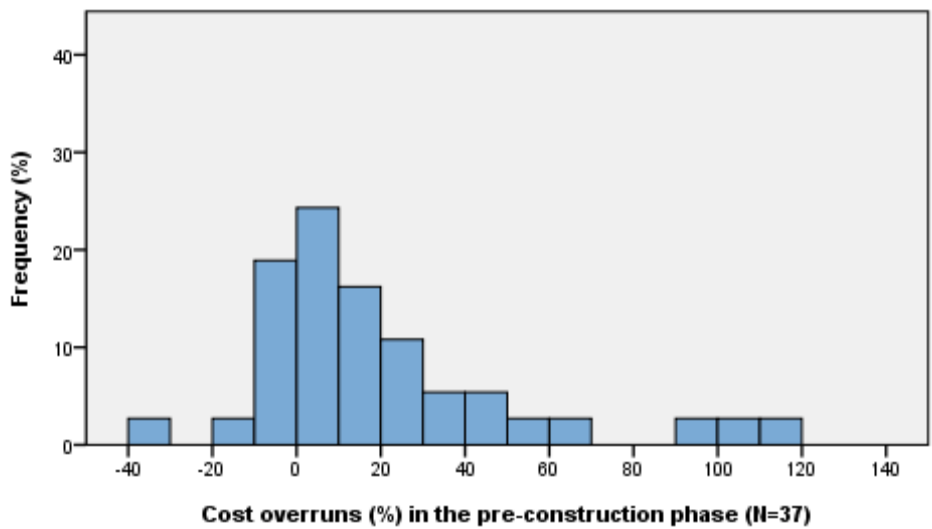

**Figure 4 Distribution of cost overruns in the pre-construction phase**

The histogram shows an asymmetric distribution around zero with a larger number of projects with cost overruns. The statistics for the cost overruns in the pre-construction phase are as follows:

- The range of cost overruns is -39.5% to 112.1%
- The average cost overrun is 19.7%
- The standard deviation is 32.6

Three projects have extremely large cost overruns between 90% and 120%. One of the projects was confronted with large delays in the development plan procedures increasing the length of the pre-construction phase and possibly the costs. Large cost overruns in the pre-construction phase suggest that the projects are relatively easy to construct and that the cost overruns are rather the result of a difficult decision-making process or large scope changes.

The main findings regarding the frequency of cost overruns in the pre-construction phase are as follows:

- In 70% of the projects, estimated costs increased, whereas in 30% of the projects, estimated costs stayed the same or decreased (p=0.020, binominal test).
- For the projects with cost overruns, the average overrun is 30.8% (SD=32.5) and for the projects with cost underruns, the average underrun is 6.5% (SD=11.3) (p=0.002, Mann-Whitney U-test).

Cost overruns are not only more frequent than cost underruns in the pre-construction phase, costs that have been underestimated are inaccurate to a larger extent than costs that have been overestimated.

Figure 5 shows the distribution of cost overruns in the construction phase.

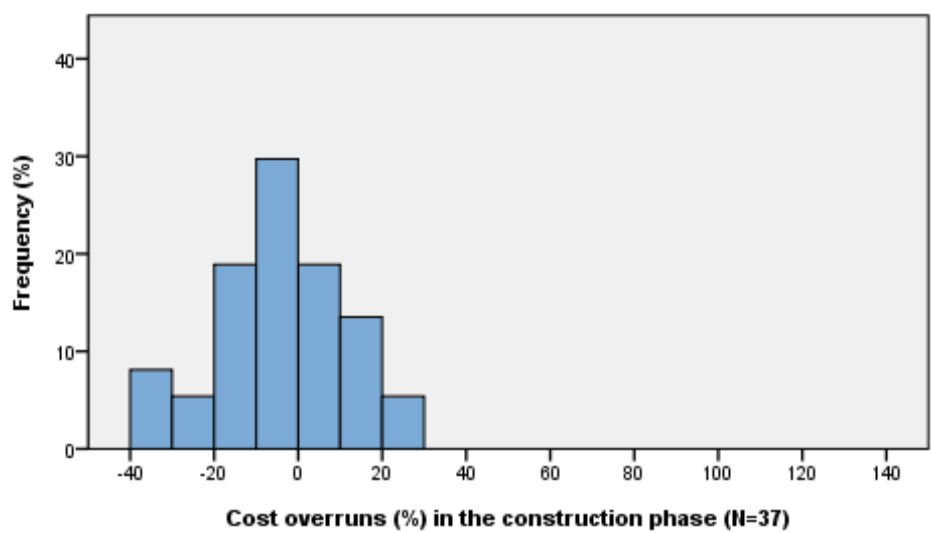

**Figure 5 Distribution of cost overruns in the construction phase**

The histogram of cost overruns in the construction phase is more symmetric than the one of the pre-construction phase but again the distribution is not symmetrically around zero; projects with cost overruns and with underruns differ. The statistics for the cost overruns in the construction phase are as follows:

- The range of cost overruns is -35.4% to 22.8%
- The average cost overrun is -4.5%
- The standard deviation is 14.4

The main findings regarding the frequency of cost overruns in the construction phase are listed below:

- In 38% of the projects cost overruns occur whereas in 62% of the projects cost underruns occur (p=0.188, binominal test).
- For the projects with cost overruns, the average overrun is 9.5% (SD=7.4) and for the projects with cost underruns, the average cost underrun is 13.1% (SD=10.4) (p=0.347, Mann-Whitney U-test).

It can be concluded that the main problem with cost overruns takes place even before construction has started. The frequency of cost overruns as well as the average overrun is larger in the pre-construction phase. The average cost overrun in the pre-construction phase is significantly higher than the average overrun in the construction phase (t=-4.118, p=0.000, paired-sample t-test).

## Conclusions and Discussion

The objective of this study was two-fold: first of all it was aimed to determine the characteristics of cost overruns in Dutch large-scale transport infrastructure projects, and secondly it was aimed to investigate whether projects are more vulnerable to cost increases in certain phases of project development.

Regarding the first objective, we found that cost overruns have been a problem for the last 20 years. Further, although in the Netherlands cost overruns are about as common as cost underruns, the average *overrun* is larger than the average *underrun*. Overall, projects have an average overrun of 16.5%.

Considering these findings we reject technical explanations as the main reason for forecasting errors in Dutch large-scale transport infrastructure projects. Technical explanations account for cost overruns in terms of imperfect forecasting techniques, inadequate data, honest mistakes, etc. If imperfect techniques were main explanations of the underestimations, an improvement in forecasting accuracy of time would be expected, since errors and their sources would be recognised and addressed through the refinement of data collection, forecasting methods, etc, but accuracy has not improved over time. The cost underestimation in Dutch projects can better be explained by psychological and political-economic explanations. The most common psychological explanation is probably "appraisal optimism". According to this explanation, promoters and forecasters are held to be overly optimistic about project outcomes in the appraisal phase, when projects are planned and decided (Fouracre et al., 1990, p. 10; Mackie & Preston, 1998; Walmsley & Pickett, 1992, p. 11; World Bank, 1994, p. 86 *in Flyvbjerg et al., 2002, p 288*). An optimistic cost estimate is a low one and if appraisal optimism is a cause of cost overrun, the actual costs would be higher than the estimated costs. Political-economic explanations see planners and promoters as deliberately and strategically underestimating costs when forecasting the outcomes of projects. They do this in order to increase the likelihood that it is their projects, and not those of the competition, that gain approval and funding (Wachs, 1989; Flyvbjerg et al. 2002). A strategic estimate of costs would be low, resulting in cost overrun. Optimism bias and strategic misrepresentation both involve deception, but where the latter is intentional - i.e., lying - the first is not. Optimism bias is self-deception.

It is expected that the percentage cost overruns that are presented in this paper and also in other studies are underestimated due to the methodology (section 2) as well as due to the use of the formal decision to build as the basis for the estimated costs. The point at which decision-makers informally decide to carry out the project is often made before the formal decision to build (Cantarelli et al., 2010). This is referred to as the *real* (informal) decision to build as opposed to the *formal* decision to build. It is highly likely that the estimated costs at the formal decision to build are larger since estimated costs usually become more accurate over time. For example, costs increased on average by 63.4% between the first estimate and the estimate at the time of the formal decision to build. Consequently, when the smaller estimated costs at the formal decision to build are used to calculate the cost overruns, the overrun will be larger. Two case studies have shown that the cost overruns were 4 to 5 times larger when the estimated costs at the informal decision to build were taken as a reference (Cantarelli et al., 2010)[12]. Because it is almost impossible to determine the real decision to build, we used the formal decision to build as a reference but recognise that the calculated cost overruns are probably higher as a result.

With respect to cost overruns during project development, the problem of cost overruns mainly occurs in the pre-construction phase, the period between the formal decision to build and the start of construction. The probability that projects incur cost overruns as well as the average overrun is higher in this phase compared to that in the construction phase. Moreover, in the construction phase, most projects involve cost underruns and the average overrun is negative.

The large difference in cost increase between the pre-construction and construction phase is remarkable. In an attempt to explain this, we have come up with four possible explanations. First of all, it could be the result of misconceived estimates. Over time, project plans become more detailed and costs can be better estimated. Secondly, the essence of the cost estimate changes over time. In the first phases of project development, the estimates are rough and have an "indicative" character whereas at the start of construction, the estimates are much more detailed and have a more "restrictive" nature that allows fewer adjustments. Thirdly, cost estimates are often optimistic and become more realistic as project plans develop. Since project plans in general change the most during the pre-construction phase, cost increases can be the result of this so-called optimism bias. Fourthly, costs could be kept low deliberately to get the project proposal accepted. After acceptance, the "real" estimates become known or scope changes are introduced (other than functional changes) that involve higher costs. Here, it is often referred to as salami-tactics, deliberately adding scope to the project.

Further, a strikingly large number of projects were identified where construction started in the same year as the decision to build. This could be the result of the methodology but it is more likely to be the result of lock-in. The informal decision to build must have been taken earlier and preparations had already taken place and procedures had been started that allowed construction to start as soon as the formal decision to build was taken.

## Areas for Further Research

This study was based on data from 78 projects and this was the best obtainable data within our research set, but further efforts to enlarge the database should be made. In addition, there are several important issues that need to be addressed in subsequent research. First of all, more insight should be obtained into the determinants of cost overruns for Dutch projects such as project type and project size. Secondly, this research has shown that projects are more prone to cost overruns in the pre-construction phase than in the construction phase. A related topic for further research is to investigate how the lengths of these phases correlate with cost overruns in the respective phases. It would further be interesting to disaggregate the cost overruns in the pre-construction and construction phase by project type and project size.

Thirdly, a systematic comparison with the Netherlands and other countries is useful to determine whether the differences between Dutch projects and those worldwide are statistically significant. We will explore these subjects, the determinants, the lengths of the project phases and the statistical significance of the difference between the Dutch and international findings in subsequent papers.

We conclude with two areas for further research, which both require additional data collection. First of all, it would be useful to consider the cost overruns for different project phases for other countries as well. This would give insight into whether it is common for projects to have the largest cost increase in the pre-construction phase or whether this is a specific feature of Dutch projects. Differences in the decision-making procedures between countries should be taken into account when drawing conclusions. Secondly, although we recognise that it is very difficult to establish the time of the informal decision to build, more research is needed into this area as it considerably influences the extent of cost overruns.

## Acknowledgement

The authors thank the Dutch Ministry of Infrastructure and the Environment, including RWS and KiM for their support in data collection. Further, the authors thank various other institutes who provided us with valuable data of projects. Special thanks to two anonymous referees for their useful comments on an earlier draft of this paper. The research was supported by the Dutch Ministry of Infrastructure and the Environment.

## Notes

1. These are the costs at the time of project completion in 2010 prices.
2. RWS, Rijkswaterstaat, is the executive agency of the Ministry of Infrastructure and the Environment, responsible for the construction and maintenance of roads and waterways.
3. The translation of the MIRT in English is based on:
   http://www.verkeerenwaterstaat.nl/english/topics/water/delta_programme/rules_and_fram ework_of_the_mirt (consulted 20-03-2010)
4. Note that the MIRT was called MIT until 2008; from 1984-1989 it was called MPP, translated here as Multi-year Passenger Transport Programme and spatial planning projects were not part of the programme, only passenger transport was included and not freight transport.

5. With the exception of MIRT 1985

6. The use of length as a surrogate criterion was verified and with the specified minimum lengths the probability of rejecting projects that should have been included is minimised.

7. The full methodological elucidation will be included in a PhD Thesis (forthcoming, 2011), of which this paper makes up a part.

8. By RWS direction Large Projects and KiM, Kennisinstituut voor Mobiliteitsbeleid (Netherlands institute for Transport Policy Analyses, an independent institute within the Ministry of Infrastructure and the Environment)

9. The GWW index concerns an index for "ground, water and road construction". The CROW is a platform of knowledge on infrastructure, traffic and transport, and public space and provides a special index for large bridges and tunnels.

10. One project had correct costs and is therefore combined with the projects with a positive cost performance record)

11. Note that the indicated cost overruns in the referred paper differ from the cost overruns presented in this paper due to the difference in the use of current and constant prices.

12. Two additional projects were excluded because their actual costs are more than €3,000 million whereas the average cost is about €95 million for the other projects.

13. The size of the project concerns here the actual costs, since these are the costs at actual opening.

# List of Figures

# List of Tables